\documentclass[aps,twocolumn,superscriptaddress,showpacs,showkeys]{revtex4}
\usepackage{graphics,graphicx,dcolumn,bm,fleqn,epic,eepic,float}
\usepackage{amssymb,amsmath,multirow,rotate,color,float}
\usepackage{times}
\definecolor{red}{rgb}{1,0,0}

\definecolor{blue}{rgb}{0,0,1}

\begin{document}
\title{A system of mobile agents to model social networks} 

\date{\today}

\author{Marta C.~Gonz\'alez}
\affiliation{Institute for Computational Physics, 
             Universit\"at Stuttgart, Pfaffenwaldring 27, 
             D-70569 Stuttgart, Germany}
\affiliation{Departamento de F\'{\i}sica, Universidade Federal do
             Cear\'a, 60451-970 Fortaleza, Brazil}
\author{Pedro G.~Lind}
\affiliation{Institute for Computational Physics, 
             Universit\"at Stuttgart, Pfaffenwaldring 27, 
             D-70569 Stuttgart, Germany}
\affiliation{Centro de F\'{\i}sica Te\'orica e Computacional, 
             Av.~Prof.~Gama Pinto 2,
             1649-003 Lisbon, Portugal}
\author{Hans J.~Herrmann}
\affiliation{Institute for Computational Physics, 
             Universit\"at Stuttgart, Pfaffenwaldring 27, 
             D-70569 Stuttgart, Germany}
\affiliation{Departamento de F\'{\i}sica, Universidade Federal do
             Cear\'a, 60451-970 Fortaleza, Brazil}

\begin{abstract}
We propose a model of mobile agents to construct social networks,
based on a system of moving particles by keeping track of the
collisions during their permanence in the system.
We reproduce not only the degree distribution, clustering coefficient
and shortest path length of a large data base of empirical 
friendship networks recently collected, but also some features 
related with their community structure.
The model is completely characterized by the
collision rate and above a critical collision rate we find the
emergence of a giant cluster in the universality class of
two-dimensional percolation. 
Moreover, we propose possible schemes to reproduce other networks of 
particular social contacts, namely sexual contacts.
\end{abstract}

\maketitle


Friendships among a group of 
people, actors working 
in the same movie or co-authors of the same paper,
are all examples of systems represented as networks, 
whose study imprinted to social networks an unquestionable
place in the field of complex networks~\cite{barabasirev,newmanrev}. 
However, the topological features of networks of acquaintances 
fundamentally differ from other networked 
systems~\cite{newmanrev,Satorras}. 
First, they are single-scale networks
and present small-world effect~\cite{Amaral}.
Second, they are divided into groups or communities~\cite{newmanrev}. 
Additionally, 
their evolution process differs from standard growth 
models as those that govern e.g.~the World Wide Web.
An interesting development in this area
is given in \cite{davidsen} where it is proposed a simple 
procedure of transitive linking to generate small-world networks.
While each one of the mentioned features can be reproduced with some previous
model, there is still no single model that 
incorporates simultaneously dynamical evolution, clustering
and community structure.  

In this Letter we show that all these characteristics can be reproduced
in a very natural way, by using standard concepts and techniques
from physical systems. 
Namely, we propose an approach to dynamical networks based on 
a system of mobile agents representing the nodes of the network. 
We will show that, due to this motion, it is possible
to reproduce the main properties~\cite{barabasirev,newmanrev}
of empirical social networks,
namely the degree distribution, the clustering coefficient (CC) and 
the shortest path length, by choosing the same average degree measured 
in the empirical networks, and adjusting only one parameter, the density 
of the system. 
The community structure emerges naturally,
without labeling {\it a priori} the community each agent 
belongs to, as in previous works~\cite{watts02}.
Moreover, this approach gives some insight to further explain
the structure of empirical networks, from a recently available 
large data set of friendship networks~\cite{addhealth} concerning 
$90118$ students, divided among $84$ schools from USA, 
constructed from an In-School questionnaire. 
The acquaintance between pairs of students was rigorously defined.
Each student was given a paper-and-pencil questionnaire and a copy
of a list with every student in the school. The student was asked 
to check if he/she participated 
in any of 5 activities with the friend: like going to (his/her) house in the
last seven days, or meeting (him/her) after school to hang out or 
go somewhere in the last seven days, etc. Other studies ~\cite{Amaral} have
used a slightly different definition of friendships 
and obtained the same kind of degree distribution, an
indication of the robustness of the concept of friendship.

Our model comprehends $N$ particles (agents) with radius $r$ 
moving continuously in a square shaped cell of linear size $L$ with periodic 
boundary conditions and low density $\rho \equiv  N/L^{2}$.
One link (acquaintance) is formed whenever two agents intercept.
After each collision, each colliding agent moves in a random direction 
with an updated velocity, till it collides again acquiring a new random 
direction, and so forth. In this way, the resulting movement alternates 
between drift (between collisions) and diffusion (collisions).
Similarly to human communities, agents arrive and depart after a certain 
time of residence, the total number of agents remaining fixed in time,
which enables the system to reach a quasi-stationary state.
Initially all agents are placed randomly,
with the same velocity modulus $v_{0}$ and random directions. 
At each time step $\Delta t$, the position $\mathbf{x}_{i}$
of agent $i$ is updated according to 
\begin{equation}
\mathbf{x}_{i}(t+1)=\mathbf{x}_{i}(t)+\mathbf{v}_{i}(t)\Delta t .
\label{position}
\end{equation}
After collisions velocity modulus of each agent, say $i$, is 
updated proportionally to its degree $k_{i}$, defined as the number of 
links connected to an agent $i$ at time $t$:
\begin{equation}
\vert \mathbf{v}_{i} (t)\vert= v_{0} + \bar{v}k_{i}(t) ,
\label{eq:velo}
\end{equation}
where $\bar{v}$ is a constant having unit of velocity and
$v_{o}$ is the initial velocity of the agents, corresponding to 
a characteristic time $\tau_{o} \equiv 1/(2 \sqrt{2\pi} r \rho
v_{o})$ between collisions.
We assume that `age' ${\cal A}_i$ is the only
intrinsic property of each agent $i$, initially randomly and homogeneously
chosen from an interval $[0,T_l]$, 
and updated as
\begin{equation}
{\cal A}_{i}(t+1)= {\cal A}_{i}(t)+\Delta t .
\end{equation}
When ${\cal A}_{i}=T_{l}$, agent $i$ leaves the system, 
all its links are removed, and a new agent replaces its position with
the initial conditions stated above, namely velocity modulus $v_{0}$
and an age randomly distributed in the range $[0,T_{l}]$.
Therefore the time of permanence of an agent in the system
is given by $T_{l}-{\cal A}_i(0)$.

\begin{figure}[htb]
\begin{center}
\includegraphics[width=7.5cm]{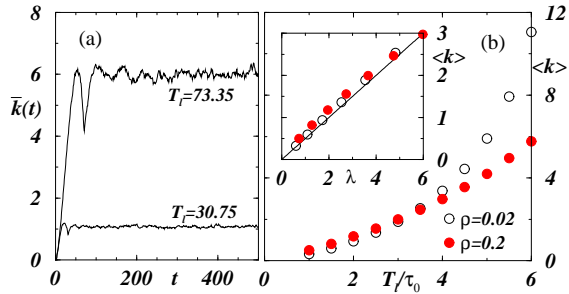}
\end{center}
\caption{\protect
   (Color online)
   {\bf (a)} Average degree $\bar{k}$ per agent as function 
             of time $t$,
             illustrating the convergence towards a QS
             state ($N=4096$).
   {\bf (b)} Average degree $\langle k \rangle $ 
             vs.~$T_l/\tau_0$ for $N=10^4$, 
             averaged over $100$ realizations.
   Inset: linear dependence between $\langle k \rangle $ and
             $\lambda$ (see text); 
             the solid line indicates $\langle k\rangle = \lambda/2$.
   In all cases $v_0=\sqrt{2}$ and $\bar{v}=1$.}
\label{fig1}
\end{figure}

After a certain transient the system reaches a quasi-stationary (QS) state.
 Thus, the degree distribution, degree correlations
and community structure depend only 
on two parameters, namely $\rho$ and $T_{l}/\tau_{o}$.
Figure \ref{fig1}a illustrates the convergence towards the QS state for
the average degree $\bar{k}(t)$ per agent.
\begin{table}[htb]
\begin{center}
\begin{tabular}{ccccc} \hline\hline
             & \mbox{\ Mean-field\ } & \mbox{\ $2D$ percolation\ } &
               \mbox{\ Mobile agents\ }   \\
$\nu$        &  $0.5$  &    $4/3\sim 1.33$    & $1.3  \pm  0.1$  \\
$\gamma$     &  $1$    &    $43/18\sim 2.39$  & $2.4  \pm  0.1$  \\
$\beta$      &  $1$    &    $5/36\sim 0.139$  & $0.13 \pm 0.01$  \\
$\sigma$     &  $0.5$  &    $36/91\sim 0.397$ & $0.40 \pm 0.01$  \\
\hline \hline
\end{tabular}
\end{center}
\caption{\protect
         Critical exponents related to the emergence of
         the giant cluster for the network of mobile agents, compared
         to the ones of mean-field and $2D$ percolation.}
\label{table1}
\end{table}

In Fig.~\ref{fig1}b we show the degree per agent $\langle k
\rangle$ vs.~$T_{l}/\tau_{o}$.
For each value of $T_{l}/\tau_{o}$ the average degree was averaged
over different snapshots in the QS regime, yielding a non-linear function 
of $T_{l}/\tau_{o}$, which depends on the chosen density.
An approximate analytical treatment of this dependence can be made and 
will be presented elsewhere. 
Further, the average degree is a function of the average
number $\lambda$ of collisions during the average residence time 
$T_{l}-\langle {\cal A} \rangle$, and is defined as
\begin{equation}
\lambda \equiv \frac{1}{v_{o}\tau_{o}}
\langle v \rangle (T_{l}-\langle {\cal A} \rangle). 
\label{eq:nc}
\end{equation}
As illustrated in the inset of Fig.~\ref{fig1}b, we find 
$\langle k \rangle = \lambda/2$ (solid line), independently of the 
density.

In the presented model, 
we find a critical value $\lambda_c=2.04$, beyond which a giant
cluster of connected nodes emerges. 
Table \ref{table1} shows the values
obtained numerically with the standard method of finite size scaling
for systems of $N=2^{10}...2^{16}$, the results are
compared with exponents for mean field and two-dimensional
percolation ($2D$). 
Since the agents move on a $2D$ plane and have only a finite life time, 
they can only establish connections within a restricted vicinity.
This effect corresponds to a connectivity which is short range
at each snapshot of the system. 
So, although our clusters are not quenched in time the 
underlying problem corresponds to short range 2D percolation.
We have also explicitly calculated the correlation length as the
linear size of clusters, and 
confirm that near the critical point this quantity diverges with 
precisely the same exponent $\nu$ obtained from the 
finite size scaling.    
\begin{figure}[t]
\begin{center}
\includegraphics*[width=7.5cm,angle=0]{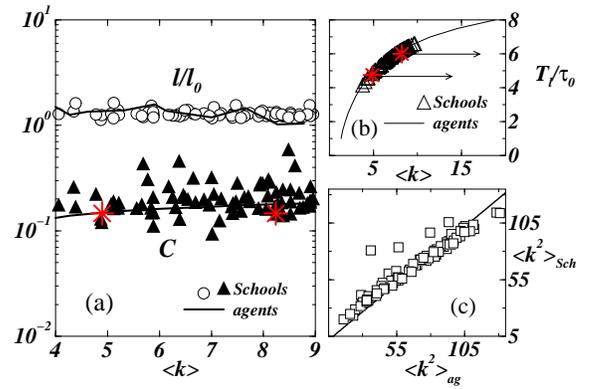}
\end{center}
\caption{\protect
         (Color online)
         {\bf (a)} Average shortest path length $l$ and clustering
                   coefficient $C$ as functions of the average degree
                   $\langle k \rangle$. Empirical data (symbols) 
                   compared to simulations (solid lines).  
         {\bf (b)} Plot of $T_{l}/\tau_{o}$ as a function of $\langle
                   k \rangle $ for the agents models (solid line). 
                   Stars illustrate two particular schools for 
                   Figs.~\ref{fig3} and \ref{fig5} having 
                   $T_{l}/\tau_{o}=4.75$ (school 1) and $6.0$ (school 2)
                   respectively.
         {\bf (c)} Second moment $\langle k^2\rangle$ for each school 
                   vs.~the second moment of the 
                   corresponding simulation with the agent model
                   (solid line has slope one).}
\label{fig2}
\end{figure}
\begin{figure}[htb]
\begin{center}
\includegraphics*[width=7.5cm,angle=0]{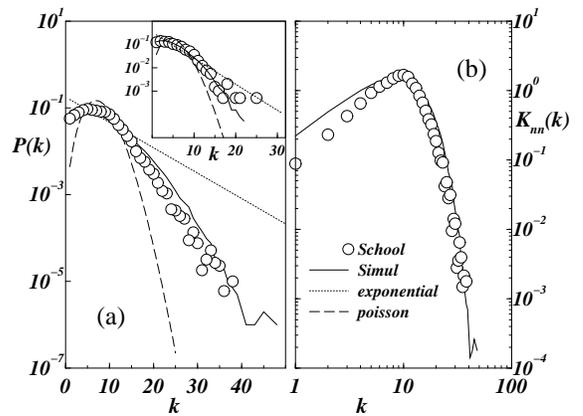}
\end{center}
\caption{\protect
        {\bf (a)} Degree distribution $P(k)$ averaged over all the schools 
        (symbols) compared to $P(k)$ of the simulations (solid line).
        The inset shows the results for a particular school (school 1).
        {\bf (b)} Average degree $K_{nn}$ of the nearest neighbors 
        as a function of $k$.
        Dashed and dotted lines indicate the Poisson and 
        exponential distributions respectively, for the same average degree
        $\langle k \rangle$.}
\label{fig3}
\end{figure}

The degree distribution $P(k)$ is a direct 
consequence of the collision rule, 
i.e.~it depends on 
$\bar{v}$ in Eq.~(\ref{eq:velo}).  
For $\bar{v}=0$, the degree distribution is well fitted 
by a Poisson distribution, 
$P_p(k)=(\langle k \rangle ^k/k!)\exp{({-\langle k \rangle})}$.
%
The degree distribution obtained 
for $\bar{v}=1$, resembles an exponential of the form 
$P_e(k)=(\langle k \rangle -1)^{-1}\exp{({-(k-1)/(\langle k \rangle -1)})}$.
However, while for small $\langle k\rangle$ the degree distribution of
the giant cluster is exponential of the form of $P_e(k)$,
for larger $\langle k \rangle $ it deviates from this shape.
The same deviation as $\langle k \rangle$ increases is in fact found in
empirical data, e.g. the friendship networks of the $84$
schools. 
For each of the schools,
Fig.~\ref{fig2}a shows the average shortest path length $l$ (circles), 
and the CC (triangles).
Solid lines indicate the results obtained for the agent model using
the same range of values of $\langle k \rangle $, averaged over
$100$ realizations with $N=2209$ and $\rho=0.1$.
Since $l$ depends on the network size, it is divided by the shortest
path length $l_{0}$ of a random graph with the same average degree
and size. 
Clearly, the agent model predicts accurately both the CC 
and the shortest path length for the same average degree.
\begin{figure}[htb]
\begin{center}
\includegraphics*[width=7.5cm,angle=0]{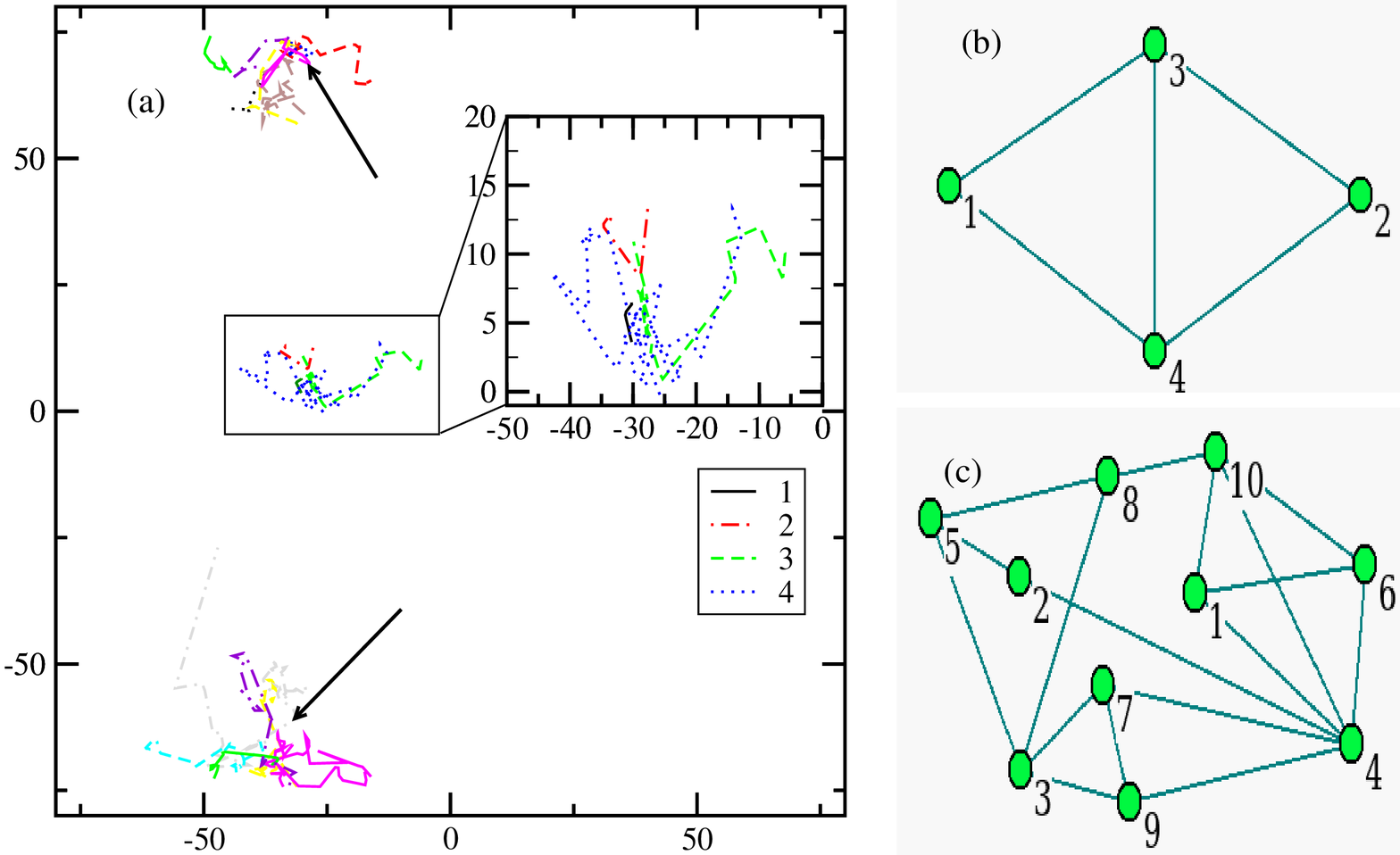}
\end{center}
\caption{\protect
        (Color online)
        {\bf (a)} Example of 
        trajectories of $4$ agents 
        (enclosed in a box and enlarged in the inset) 
        and $10$ agents (showed by arrows) forming 
        a $3$-clique sketched in {\bf (b)} and {\bf (c)}.}
\label{fig4}
\end{figure}
\begin{figure}[htb]
\begin{center}
\includegraphics*[width=7.5cm,angle=0]{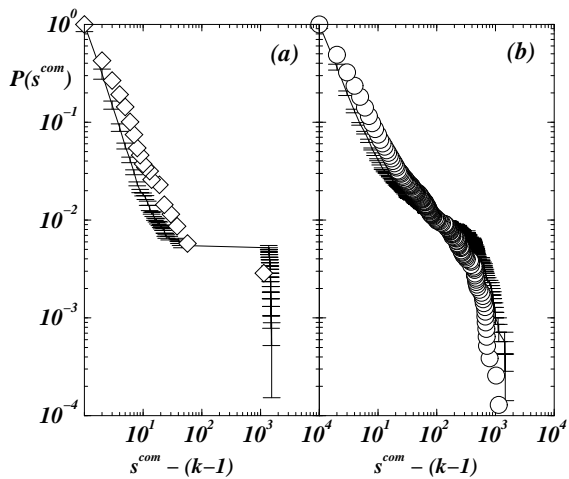}
\end{center}
\caption{\protect
        {\bf (a)} Distribution of community size $s$ of $3$-clique 
        communities for one particular school (school 2)
        {\bf (b)} the corresponding average over the $84$ schools of the
        data set. Empirical data (symbols) compared to simulations (solid
        lines with error bars).}
\label{fig5}
\end{figure}
\begin{figure}[htb]
\begin{center}
\includegraphics*[width=7.0cm,angle=0]{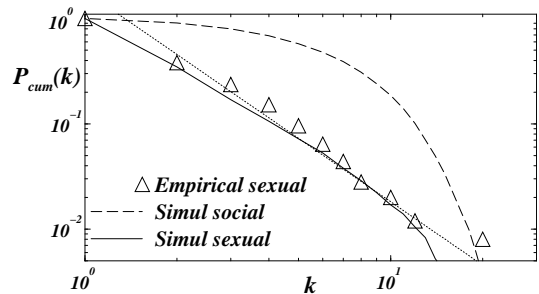}
\end{center}
\caption{\protect
         Cumulative degree distribution of the number $k$ of sexual
         partners in a real empirical network of sexual contacts
         (triangles) with $250$ individuals, compared with the simulation 
         of the agent model
         (solid line), the dotted line is a guide to the eye with
         slope $2$.
         Here $N=4096$, $T_l/\tau_{o}=5.5$ and $\langle k \rangle=7.32$
         and the average size of the resulting sexual network is $220$.}
\label{fig6}
\end{figure}

By computing the average degree $\langle k \rangle$ of each
school one is able to obtain the value of $T_{l}/\tau_{o}$
for which the agent model reproduces properly the empirical
data, as illustrated in Fig.~\ref{fig2}b.
Here solid lines indicate the prediction curve for the agent model,
while triangles indicate the values of $T_{l}/\tau_{o}$ chosen to
reproduce the social network of the schools with the resulting
value of $\langle k\rangle$. 
Moreover, the second moment $\langle k^2\rangle_{ag}$ obtained 
with the simulations of the agent model is a rescaling of the same
quantity $\langle k^2\rangle_{Sch}$ measured for the empirical school
networks, as shown in Fig.~\ref{fig2}c.
  
Figure \ref{fig3}a shows the degree distribution averaged over
all the schools, compared with the average of the ones obtained from 
the agent model simulations using the chosen values of $T_l$ according 
to the relation sketched in Fig.~\ref{fig2}b.
As one clearly sees, the degree distribution obtained with the 
agent model fits much better the empirical data than the exponential
(dotted line) or Poisson (dashed line) distributions for a
given $\langle k \rangle$. The inset in the figure~\ref{fig3}a
shows the comparison of the network of one particular school
(school 1 in Fig.~\ref{fig2}), and
the average over 20 realizations of its corresponding model
(with $T_{l}/\tau_{o}=4.75$).
  
Degree correlations can be quantified by computing $K_{nn}(k)$, 
the average degree of the nearest neighbors
of a vertex of degree $k$ \cite{Satorras}. Figure \ref{fig3}b
shows a good agreement of this value between real data and model for
the same networks of Fig.~\ref{fig3}a. Similar to 
other social networks the mixing is 
assortative~\cite{newmanrev}, i.e. 
$K_{nn}$ increases with k, but in contrast to networks
with scale free degree distribution 
(i.e. collaboration networks), $K_{nn}(k)$ for friendship networks
present a cutoff due to the rapid 
decay in the degree distribution.

Further, the typical community structure found in social networks,
, is also reproduced with the agent model.
Here, we use a precise definition of network community recently 
proposed~\cite{Palla} based on the concept of $k$-clique community.
In Fig.~\ref{fig4} we plot the system of mobile agents, drawing only the
trajectories of the agents which belong to two $3$-clique communities,
having $4$ and $10$ agents and sketched in
Fig.~\ref{fig4}b and Fig.~\ref{fig4}c respectively.
Agents that form a community share a region in space and 
agents with larger trajectories are responsible
for building up the community. 
It should be pointed out that the agent motion in the
system has not the straightforward meaning of human motion
in physical space, but may be better related with affinities
among in\-di\-vi\-duals.

Figure \ref{fig5}a shows the size distribution of $3$-clique 
communities in a particular school (school 2) compared with
the simulation for the suitable value of $T_{l}/\tau_0$ (see Fig.~\ref{fig2}),
while in Fig.~\ref{fig5}b the average over all schools is compared with 
the average over $10$ realization of the corresponding model for each school. 
In both cases, the agent model reproduces the distribution of community size
observed for the empirical data, particularly the feature related with
the existence of a big community having a large fraction of the population,
namely $s \sim 10^{3}$ agents.

In the particular case of sexual contacts it has been reported
that the degree distribution presents a power-law~\cite{Liljeros}.
Figure \ref{fig6} shows with triangles the cumulative degree
distribution  of a sexual contact network extracted from a tracing
study for HIV tests in Colorado Springs (USA) with $250$
individuals~\cite{cospring2}.
The dashed line indicates the degree distribution of a social contact
network simulated with the agent model while the solid line is the
degree distribution of a subset of contacts from the social network.
The contacts in the subset are chosen by assigning to each agent
an intrinsic property which enables one to select from all 
the social contacts the ones which are sexual. 
Namely, when two agents form a link, as stated before, this link is 
now marked as a 'sexual contact' if the sum of the property values 
of the two agents is greater than a given threshold.
These property values are assigned to the agents 
with an exponential distribution and the conditional
threshold is $\ln{N}/2$, following the scheme 
of intrinsic fitness proposed in another context by Caldarelli
et. al.~\cite{Caldarelli}. Interestingly, one is able to extract 
from the typical distributions of social contacts shown throughout the
paper, power-law distributions in QS which resemble much
the ones observed in real networks of sexual contacts.

In conclusion, we presented a novel approach to construct contact
networks, based on a system of mobile agents.
For a suitable collision rule and aging scheme we have shown that
one is able to produce quasi-stationary states which reproduce
accurately the main statistical and topological features observed in
recent empirical social networks. 
The QS state of the agent model is fully characterized by one single
parameter and yields a phase transition belonging to the 
universality class of two-dimensional percolation. 
Moreover, we showed that, by introducing an additional property
labeling the ability to select a particular type of social contact,
e.g.~sexual contacts, the degree distributions reduce to power-law
distributions as observed in real sexual networks.
Summarizing, 
we gave evidence that 
motion of the nodes is a fundamental feature to reproduce social networks,
and therefore the above model could be important to improve 
the study and may serve as a novel approach to model empirical 
contact networks.

The authors would like to thank J.~K\'ertesz, J.S.~Andrade and
M. Barth\'el\'emy for useful discussions. 
MCG thanks 
DAAD (Germany) and PGL thanks 
FCT (Portugal) for financial support. 

\end{document}